\documentclass[10pt,a4paper]{article}
\usepackage[dvips]{graphicx}
\usepackage{amssymb}
\usepackage[left=3cm,top=3cm,bottom=3cm]{geometry}
\usepackage{setspace}

\begin{document}

\noindent This manuscript version is distributed under the CC-BY-NC-ND (Creative Commons) license.\\

\noindent It has appeared as:\\
Van Albada SJ, Robinson PA. Transformation of arbitrary distributions to the normal distribution with application to EEG test-retest reliability (2007) J
Neurosci Methods 161: 205--211, DOI: 10.1016/j.jneumeth.2006.11.004\\

\newpage

\title{Transformation of arbitrary distributions to the normal distribution with application to EEG test-retest reliability}
\author{S. J. van Albada$^{a,b}$ \and P. A. Robinson$^{a,b,c}$}
\maketitle

\begin{center}
$^a$School of Physics, The University of Sydney\\
New South Wales 2006, Australia\\
$^b$The Brain Dynamics Centre, Westmead Millennium Institute\\
Westmead Hospital and Western Clinical School of the University of Sydney\\
Westmead, New South Wales 2145, Australia\\
$^c$Faculty of Medicine, The University of Sydney\\
New South Wales 2006, Australia\\


\end{center}


\begin{abstract}

Many variables in the social, physical, and biosciences, including neuroscience, are non-normally distributed. To improve the statistical properties of such data, or to allow parametric testing, logarithmic or logit transformations are often used. Box-Cox transformations or ad hoc methods are sometimes used for parameters for which no transformation is known to approximate normality. However, these methods do not always give good agreement with the Gaussian. A transformation is discussed that maps probability distributions as closely as possible to the normal distribution, with exact agreement for continuous distributions. To illustrate, the transformation is applied to a theoretical distribution, and to quantitative electroencephalographic (qEEG) measures from repeat recordings of 32 subjects which are highly non-normal. Agreement with the Gaussian was better than using logarithmic, logit, or Box-Cox transformations. Since normal data have previously been shown to have better test-retest reliability than non-normal data under fairly general circumstances, the implications of our transformation for the test-retest reliability of parameters were investigated. Reliability was shown to improve with the transformation, where the improvement was comparable to that using Box-Cox. An advantage of the general transformation is that it does not require laborious optimization over a range of parameters or a case-specific choice of form.
\end{abstract}

\section{Introduction}
The purpose of this paper is to bring to the attention of neuroscientists a general-purpose method of transforming non-normal distributions to the Gaussian. Remarkably, although the method is quite simple and its subcomponents are even mentioned in some textbooks, it is not known in the relevant applied scientific literature, while enormous effort is devoted to finding ad hoc or approximate solutions. Some of the advantages of the method described here lie in its ease of application (an identical recipe can be followed for \emph{any} data), the rendering of the data into a familiar form, and the convenience of interpretation this affords. 

Although non-normally distributed quantitative electroencephalographic (qEEG) parameters are used as a test case here, there are many examples of neuroscientific variables that follow skewed or kurtotic distributions. Among them are hippocampal and ventricular volumes (Lloyd et al., 2004), age of onset of clinical conditions (see, e.g., Bellivier et al., 2003), fitted parameters for a model of changes in neurotransmitter concentrations (Napper, Pianta, and Kallionatis, 2001), questionnaire responses, and many more. Further application of the method is found in the study of test-retest reliability, since Dunlap et al. (1994) showed that skew and kurtosis reduce test-retest correlations, and potentially in the transformation of non-normally distributed variables in data mining (Ultsch, 2000).

The existing literature reveals a wide interest in normalizing transformations. These are relevant in many disciplines of engineering and the social, physical, and biosciences, but qEEG is used as an example here. In studies of qEEG, it is common to apply the logarithmic transformation to improve the normality properties for absolute power, and to apply the logit transformation $(\log[x/(1-x)])$ for relative power (John et al., 1980), where log represents the natural logarithm, and $x$ is a dimensionless version of the quantity being transformed. Gasser et al. (1982) investigated the transformations log($x$), log($x+1$), $\sqrt{x}$, $\sqrt[3]{x}$, 1/$\sqrt{x}$ for absolute power, and arcsin $\sqrt{x}$ and log[$x/(1-x)$] for relative power. They found log($x$) to be the best transformation for the absolute power except in the delta band, where $\sqrt{x}$ may be appropriate, and confirmed the superiority of \mbox{log[$x/(1-x)$]} over other transformations for relative power. Oken and Chiappa (1988) investigated the reduction in skewness resulting from square root and logarithmic transformations of EEG power measures and the logit transformation for relative powers. They anticipated a decrease in the coefficient of variation (standard deviation divided by the mean) of spectral measures on transformation to normality, similar to the results of Dunlap et al. (1994).

The statistical literature describes a range of more systematic methods for transforming distributions toward the Gaussian. 
Notably, Box and Cox (1964) considered a family of transformations that result to a good approximation in data with normally distributed, homoscedastic (constant variance) errors, and a simple relation to the predictor variables. These transformations depend either on a single parameter $\lambda$, or on two parameters $\lambda = (\lambda_1, \lambda_2)$, and take the form
\begin{equation}
y^{(\lambda)} = \left\{ \begin{array}{ll}
\frac{y^{\lambda}-1}{\lambda} & (\lambda \neq 0),\\
\textrm{log}~ y & (\lambda = 0),
\end{array} \right. 
\end{equation}
for $y > 0$, or
\begin{equation}
y^{(\lambda)} = \left\{ \begin{array}{ll}
 \frac{(y+\lambda_2)^{\lambda_1}-1}{\lambda_1} & (\lambda_1 \neq 0),\\
\textrm{log}~ (y+\lambda_2) & (\lambda_1 = 0),
\end{array} \right.
\end{equation}
for $y > -\lambda_2$. The parameters can be estimated using either maximum likelihood or Bayesian methods, under the assumptions that for each value of the predictor variables, $y^{(\lambda)}$ is normally distributed with constant variance, and that the observations can be described by a linear model. This leads to values for $\lambda_1$ and $\lambda_2$ which minimize the residual sum of squares in the analysis of variance of 
\begin{equation}
\frac{(y+\lambda_2)^{\lambda_1}-1}{\lambda_1[\mathrm{gm}(y+\lambda_2)]^{\lambda_1-1}},
\end{equation}
where $\mathrm{gm}$ is the geometric mean, and $\lambda_2$ may be taken to be zero. From the confidence interval(s) for the parameter(s) one can pick a value that facilitates interpretation of the transformed variable. While the Box-Cox transformation is aimed at obtaining normally distributed errors relative to a linear regression line, the method has been adapted to find transformations that result to a good approximation in normality of the sample as a whole. The method remains approximate, since only a limited class of transformations is considered. 

Other examples from a vast literature on normality transformations include Draper (1952), who discussed three families of transformations and methods for finding the relevant coefficients, and Chen and Tung (2003), who described several variants of a third-order polynomial transformation. Each method has its drawbacks: determining the coefficients is cumbersome, the transformations are often only one-to-one in certain regimes, and even then achieve only approximate normality. The wide interest in normalizing transformations is also evidenced by the highly cited paper of Gasser et al. (1982). To spare researchers and practitioners the effort of applying complicated transformations that can be different in form for each new type of data, and only render distributions approximately normal, we discuss here the simple general solution.

In Sec.~2 a transformation is discussed that brings distributions as close as possible to the Gaussian, and takes a reasonably simple form. Sec.~3 illustrates the use of the transformation on an empirical set of non-normal EEG spectral measures. The implications of the transformation for the test-retest reliability of the quantities is discussed.

\section{Methods}

In Sec.~2.1 the form of the transformation is derived. Sec.~2.2 deals with its application to empirical distributions, which are not continuously defined. Sec.~2.3 gives a short description of the data set and the statistical methods used to test normality and test-retest reliability.  

\subsection{The Transformation}

The transformation consists of two parts, both of which are founded on the fundamental law of probabilities, which states that the infinitesimal area under a probability density function (PDF) is invariant under one-to-one transformations \mbox{$x \mapsto y(x)$}; i.e.,
\begin{equation}\label{fund}
|p_y(y) dy| = |p_x(x) dx|.
\end{equation}
Here $p_y$ and $p_x$ are probability densities, and $dx$ and $dy$ are infinitesimal increments. Equation (\ref{fund}) suggests a way to transform a continuous variable into a variable that is uniformly distributed, as follows. Substituting the PDF for the uniform distribution, $p_x(x)=1$, we have
\begin{equation}
p_y(y) = \left| \frac{dx}{dy}\right|.
\end{equation}
The solution of (5) is $x = \pm P_y(y)$, where $P_y(y)$ is the indefinite integral of $p_y(y)$. In other words, $p[P_y(y)]=p_x(\pm x)=1$, showing that the cumulative distribution function (CDF) of a continuous variable is itself uniformly distributed on the interval (0,1). This fact was noted by L{\'e}vy (1937) and Rosenblatt (1952), and is now a textbook result.

All we now need to do is find a transformation from a uniform deviate on (0,1) to a normally distributed quantity. If we denote the standard normal PDF by $\phi(y)$, we have to solve $dx/dy=\phi(y)$. As before, we can integrate to obtain $x= \Phi(y)$, or, since the CDF is always monotonically increasing and hence invertible, $y = \Phi^{-1}(x)$. The inverse CDF of the standard normal distribution is
\begin{equation}\label{transf1}
\Phi^{-1}(x) = \sqrt{2}~ \mathrm{erf}^{-1}(2 x -1).
\end{equation}
where erf represents the error function. The variable $y = \Phi^{-1}(x)$ will follow the normal distribution if $x$ is itself uniformly distributed. But we saw above that a uniform distribution can be obtained from any variable by taking the CDF. Therefore, if we start from a variable $v$ following a continuous PDF, the transformed variable
\begin{equation}\label{transf2}
y(v) = \sqrt{2} ~\mathrm{erf}^{-1}[2 P_v(v) - 1]
\end{equation}
has a normal distribution with mean $\mu = 0$ and variance $\sigma^2 = 1$. The mean and standard deviation can be adjusted by multiplying by the desired standard devation and adding a constant, which results in
\begin{equation}\label{withmean}
y(v) = \mu + \sigma \sqrt{2} ~\mathrm{erf}^{-1}[2 P_v(v) - 1].
\end{equation}
The transformation (\ref{withmean}) is order-preserving.

Not surprisingly, a thorough literature search reveals that the transformation (\ref{withmean}) has been known among statisticians for some time, and is for instance mentioned in Liu and Der Kiureghian (1986). However, its use appears to have been almost entirely restricted to the technical statistics literature, and it is essentially unknown to applied workers who could benefit most from the method.

The function (\ref{transf1}) proves to be the inverse of the first of a set of functions proposed by Rosenblatt (1952) to map the multivariate normal distribution onto the multidimensional uniform distribution. For instance, in the two-dimensional case, letting $\Phi(z) = (2 \pi)^{-1/2} \int^z_{-\infty} e^{-t^2/2}dt$, the transformations of normally distributed variables $y_1, y_2$ with means $\mu_1,\mu_2$, variances $\sigma_1^2, \sigma_2^2$ and correlation coefficient $\rho$, to uniform deviates $x_1, x_2$ on $(0,1)\times(0,1)$ are (Rosenblatt, 1952)
\begin{eqnarray}\label{two-d}
x_1 &=& P(y_1) = \Phi\left(\frac{y_1 - \mu_1}{\sigma_1}\right),\\
x_2 &=& P(y_2|y_1) = \Phi \left(\frac{y_2 - \mu_2 + \frac{\rho \sigma_1}{\sigma_2}(y_1 - \mu_1)}{\sigma_2 \sqrt{1-\rho^2}} \right).
\end{eqnarray}
The inverse of (9) and (10) is given by
\begin{eqnarray}\label{inverse}
y_1 &=& \sigma_1 \Phi^{-1}(x_1) + \mu_1, \\
y_2 &=& \sigma_2 \sqrt{1-\rho^2} ~\Phi^{-1}(x_2) + \mu_2 - \frac{\rho \sigma_1}{\sigma_2}(y_1-\mu_1). 
\end{eqnarray}
In particular, when the quantities $y_1$ and $y_2$ are uncorrelated, (\ref{inverse}) factorizes into two separate realizations of the transformation (\ref{withmean}). However, note that in general $y_1$ and $y_2$ will not have the given correlation or even be normal unless $x_1$ and $x_2$ obey (9) and (10).

If the CDF can be determined analytically, applying the transformation (\ref{transf2}) is straightforward. As an example to show how well this works, we consider the two-parameter Weibull distribution, and choose parameters for which the distribution deviates strongly from normality, with nonzero mean and skew, and excess kurtosis. The density and distribution functions are given by
\begin{eqnarray}\label{Weibull}
p(x) &=& ab^{-a}x^{a-1}e^{-(x/b)^a},\\
P(x) &=& 1-e^{-(x/b)^a},
\end{eqnarray}
defined on $x \in [0, \infty)$. The transformation to the standard normal distribution is
\begin{equation}\label{weibulltransf}
y(x) = \sqrt{2} ~\mathrm{erf}^{-1}\left[1-2 e^{-(x/b)^a} \right].
\end{equation}
The skewed and leptokurtic distribution (\ref{Weibull}) with shape parameter $a = 9$ and scale parameter \mbox{$b = 3$} is shown in Fig.~1, along with the standard normal distribution obtained after the transformation (\ref{weibulltransf}).

\begin{figure}[!ht]
\centering
  \includegraphics[width=200pt, height=200pt, angle=270]{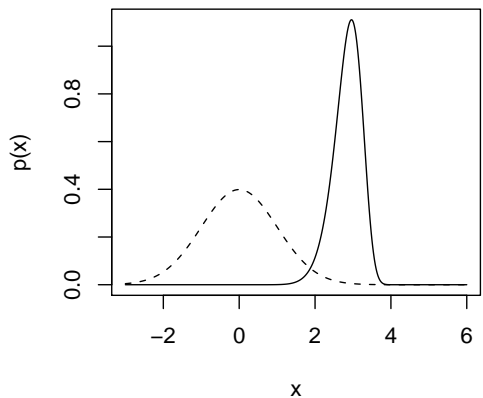}
  \caption{The Weibull PDF with shape parameter $a=9$ and scale parameter $b=1$ (solid) and the standard normal PDF after transformation (dashed).}
\end{figure}

One case of interest is the lognormal distribution, for which the transformation reduces to taking the logarithm, which is already widely used in practice.

\subsection{Use on Empirical Distributions}

In Sec.~3.1 we apply the transformation (7) to empirical distributions rather than to continuously defined analytic distributions. When working with empirical distributions, we compute the empirical distribution function (EDF) instead of an analytic CDF. For brevity only continuous variables are considered here.
The standard definition of the EDF is
\begin{equation}\label{EDF}
F_N(x) = \frac{1}{N}\sum_{i=1}^N I(x_i \leq x),
\end{equation}
where $N$ is the number of observations in the sample, and $I$ is an indicator function having a value of $1$ if the argument is true, and $0$ otherwise. In other words, (\ref{EDF}) gives the fraction of observations smaller than or equal to $x$. However, we do not want to map the largest observation to the value $1$, nor the smallest to 0, since the extremes of an underlying continuous distribution are never sampled in an empirical study. Therefore we subtract $1/(2N)$ from the normalized rank numbers computed in (\ref{EDF}), obtaining an EDF that runs from $1/(2N)$ for the smallest observation, in steps of $1/N$, to $1-1/(2N)$ for the largest observation. 

The set of values representing the EDF is uniformly distributed. We then apply the transformation (\ref{transf1}) to obtain a distribution that is symmetric and as close as possible to the Gaussian. This can be done using any mathematics package that provides the inverse error function (e.g., Matlab).

\subsection{EEG Example}

As an empirical example, we apply the transformation (7) to a number of qEEG measures obtained from repeat Cz-electrode recordings of 32 subjects in Sec.~3.1. We stress that these data constitute an illustrative example, and use of the method is \emph{not} restricted to EEG measures. 

The recordings were done at Westmead Hospital in Sydney and at Queen Elizabeth Hospital in Adelaide in the context of a reproducibility study of the EEG. All subjects were healthy males with an age at the first session in the range of 18 to 27 years (mean = 22.3, sd = 2.7). Eyes-closed resting scalp EEG was recorded with a NuAmps amplifier (Neuroscan) at 26 electrode sites according to the 10-20 international system. The sampling rate was 500 Hz and a linked-ears reference was used. Data were low-pass filtered at 40 dB per decade above 100 Hz after correcting off-line for eye movements. Power spectra were computed from 2 minutes of EEG by multiplying sequential 2.048 s epochs by a Welch window and performing a Fast Fourier Transform with 0.49 Hz resolution. The intervals between recordings ranged from 1 week to several months. Measures include powers in five frequency bands (delta, theta, alpha, beta, gamma), total power, relative band powers (defined as the power in a single band divided by the total power), and spectral entropy. Band power limits for the study were as follows: delta \mbox{(0.2--3.7 Hz)}, theta \mbox{(3.7--8.1 Hz)}, alpha \mbox{(8.1--12.9 Hz)}, beta \mbox{(12.9--30.5 Hz)}, and gamma \mbox{(30.5--49.6 Hz)}. 

The Anderson-Darling (AD) test for normality (see for instance Thode, 2002) is used to determine the degree of correspondence with the normal distribution before and after transformation. The results are also compared with the logarithmic transformation for band powers and total power, the logit ($\log[x/(1-x)]$) transformation for relative powers, and the Box-Cox transformation for all spectral measures. The latter was implemented by maximizing the p-value of the AD-test over a range of values for $\lambda_1$ and $\lambda_2$. 

The test-retest reliabilities of the spectral measures, quantified by Pearson correlations, were determined over the first six weeks of recording without transformation and after applying the various transformations. A paired-samples \emph{t}-test was used to compare the average correlation before and after transformation to normality. Finally, the reliabilities were compared with those computed using nonparametric Spearman correlations.

\section{Results}

To illustrate the use of the transformation, we apply it to the empirical distributions of a number of qEEG measures in Sec.~3.1 and show that it achieves optimal agreement with the normal distribution. Sec.~3.2 contains a further illustrative application to a study of qEEG test-retest reliability. 

\subsection{Empirical Example: EEG Spectral Data}

Histograms of the five sets of band powers are presented in the left column of \mbox{Fig. 2.} All distributions are greatly skewed to the right. Taking the logarithm brings the distributions closer to normal, but some skew remains in the delta band, while skew is overcorrected in the alpha band (middle column of Fig.~2).

\begin{figure}[!ht]
\centering
  \includegraphics[width=450pt, height=350pt,angle=270]{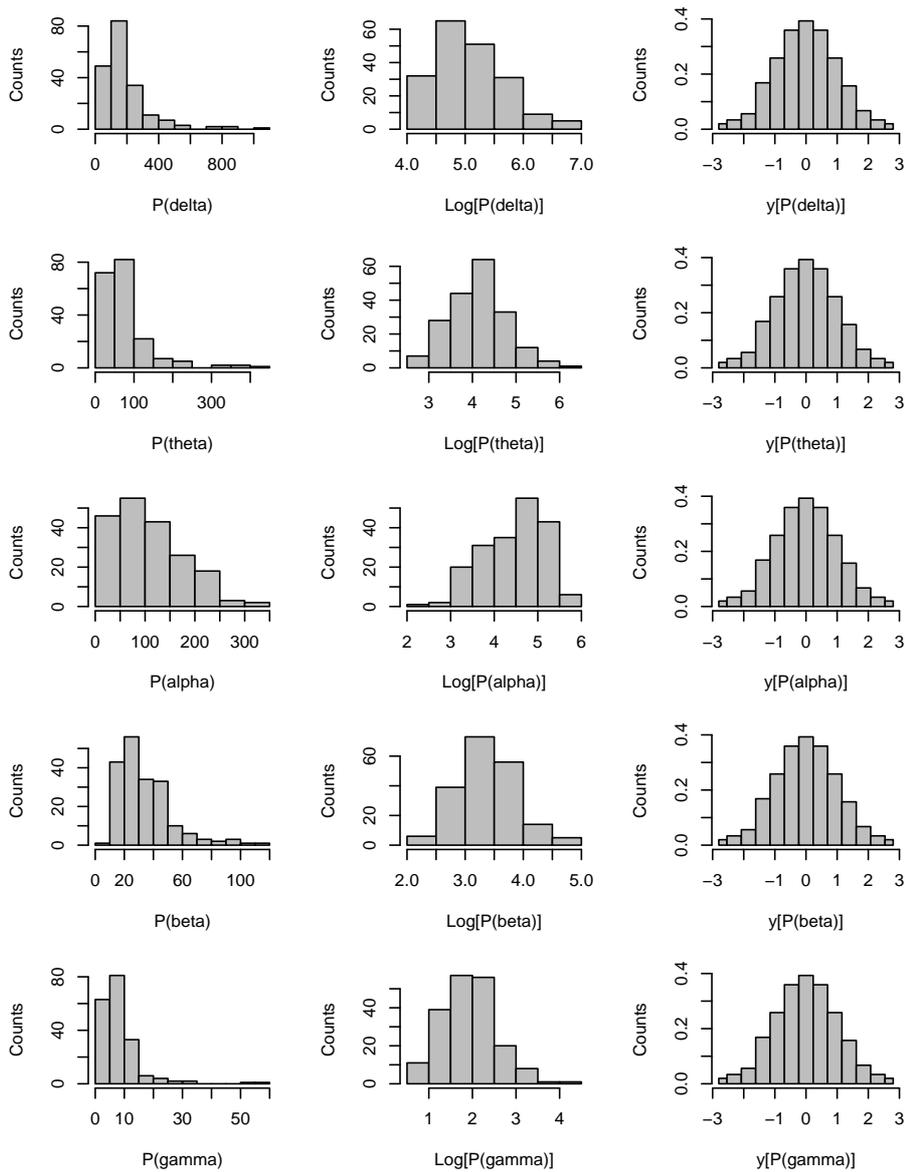}
  \caption{Histograms for band powers (indicated by P) corresponding to 193 eyes-closed EEGs of 32 healthy subjects recorded at the Cz electrode in a longitudinal study, giving the number of observations in each bin. Untransformed quantities are plotted in the left column, log-transformed data in the middle column, and data transformed with (\ref{transf2}) in the right column. The symbol y represents the transformation function.}
\end{figure}

Significance values for the Anderson-Darling (AD) test for normality of band power data from this study, before and after taking the logarithm, are listed in Table~1. It is seen that, although taking the logarithm improves the normality of the data, agreement with the normal distribution is still poor especially for the alpha and delta bands. Results of the AD-test are also listed for total power, spectral entropy, and relative band powers before and after a logit transformation. Relative band powers are defined as the corresponding band power divided by the total power and indicated by `rel'. None of the untransformed relative band powers, and only three of the relative band powers transformed with the logit transformation, pass the normality test at the $0.05$ significance level. No result is listed for spectral entropy, since it does not have a commonly used normalizing transformation.

\begin{table}[ht]
\begin{tabular}[ht]{ll|ll|ll|ll}
\hline
\textbf{Quantity} & \textbf{p-value} & \textbf{Quantity} & \textbf{p-value} & \textbf{Quantity} & \textbf{p-value} & \textbf{Quantity} & \textbf{p-value} \\
\hline 
delta power & $<$ 2.2e$-$16 & log delta & 0.0039 & delta$_{BC}$ & 0.99 & delta$_T$ & 1 \\
theta power & $<$ 2.2e$-$16 & log theta & 0.19 & theta$_{BC}$ & 0.25 & theta$_T$ & 1 \\
alpha power & 6.6e$-$09 & log alpha & 0.00061 & alpha$_{BC}$ & 0.056 & alpha$_T$ & 1 \\
beta power & 1.2e$-$15 & log beta & 0.34 & beta$_{BC}$& 0.61 & beta$_T$ & 1 \\
gamma power & $<$ 2.2e$-$16 & log gamma & 0.20 & gamma$_{BC}$ & 0.53 & gamma$_T$ & 1 \\
total power & 1.0e$-$10 & log total & 0.12 & total$_{BC}$ & 0.26 & total$_T$ & 1 \\
entropy & 3.6e$-$08 &&& entropy$_{BC}$& 0.082 &entropy$_{T}$ & 1\\
\hline
rel delta & 0.083 & logit rel delta & 0.34 &rel delta$_{BC}$& 0.71& rel delta$_T$ & 1 \\
rel theta & 1.5e$-$08 & logit rel theta & 0.69 &rel theta$_{BC}$& 0.95 & rel theta$_T$ & 1 \\
rel alpha & 0.0030 & logit rel alpha & 0.0016 &rel alpha$_{BC}$& 0.0080 & rel alpha$_T$ & 1 \\
rel beta & 2.4e$-$06 & logit rel beta & 0.31 &rel beta$_{BC}$& 0.34& rel beta$_T$ & 1 \\
rel gamma & $<$ 2.2e$-$16 & logit rel gamma & 0.0024 &rel gamma$_{BC}$ & 0.37& rel gamma$_T$ & 1 \\
\hline
\end{tabular}
\caption{Significance values for the Anderson-Darling test for normality applied to qEEG measures derived from 193 eyes-closed EEGs of 32 healthy subjects recorded at the Cz electrode in a longitudinal study. Absolute and relative band powers are tested before and after applying the commonly used transformations toward normality (logarithmic for absolute band powers; logit ($\log[x/(1-x)]$) for relative band powers). The second column is left blank for spectral entropy, for which no widely used transformation toward the normal distribution is known. All quantities are also tested after the Box-Cox transformation (3), indicated by a subscript $BC$, and after the transformation (7), indicated by a subscript $T$. The latter is seen to yield optimal agreement with the normal distribution.}
\end{table}

Better results are obtained using the Box-Cox transformation, although relative alpha power still does not pass the normality test at the 0.05 significance level. The parameters $\lambda_1$ and $\lambda_2$ used in the transformations are listed in Table~2. For band powers and total power, the optimal values for $\lambda_1$ are all close to zero, confirming that the logarithmic transformation can improve the normality of these measures. However, better agreement with the normal distribution is reached when adding a constant ($\lambda_2$) before taking the logarithm. Judging from Table~1, Box-Cox performed slightly better than the logit transformation, and much better for relative gamma power.

\begin{table}[ht]
\begin{tabular}[ht]{lll|lll}
\hline
\textbf{Quantity} & $\lambda_1$ & $\lambda_2$ &  \textbf{Quantity} & $\lambda_1$ & $\lambda_2$  \\
\hline 
delta power & 0.0 & $-$42.2 &   rel delta &   0.55 & $-$0.1  \\
theta power & $-$0.35 & 18.3 &  rel theta &  $-$0.25  & 0.1  \\
alpha power & 0.35  & $-$10.2 &   rel alpha &    0.75 & 0.0 \\
beta power &  $-$0.15  & $-$0.6 &  rel beta &  0.0    & 0.0 \\
gamma  power &   0.0  & $-$0.9 & rel gamma & $-$0.29 & 0.0  \\
total power & 0.35 & $-$100.0 &  && \\
entropy & 2.75 & $-$0.30 & && \\
\hline
\end{tabular}
\caption{Coefficients for the Box-Cox transformations (3) leading to optimal agreement with the normal distribution.}
\end{table}

In terms of normality, the transformation (7) does better than all other transformations; with this method, both the KS-test (see for instance Conover, 1971) and the AD-test for normality yield a p-value of 1 for all spectral measures.

 The histograms resulting after applying (7) to band powers are plotted in the right column of Fig.~2. Note that, although all sets of band powers have the same distribution after transformation, differences between individuals are preserved because they are located in different parts of the distributions. 

\subsection{Test-Retest Reliability}

One application of the methods outlined above is to test-retest reliability, which has been shown to improve with the normality of data when the original variables are highly skewed (Dunlap et al., 1994). Table~3 lists the average Pearson correlation coefficients for band powers, total power, and spectral entropy over the first six weeks of testing, with and without the relevant transformations. For comparison, Spearman rank correlations are also given.

\begin{table}[ht]
\begin{tabular}[ht]{llllll}
\hline
\textbf{Quantity} & \textbf{r (original)} & \textbf{r (log/logit)} & \textbf{r (Box-Cox)} & \textbf{r (normal)} & $\rho$\\
\hline 
delta power & 0.36 & 0.47 & 0.51 & 0.51 & 0.45\\
theta power & 0.83 & 0.87 & 0.87 & 0.85 & 0.83\\
alpha power & 0.85 & 0.88 & 0.88 & 0.88 & 0.87\\
beta power & 0.73 & 0.78 & 0.79 & 0.78 & 0.77\\
gamma power & 0.45 & 0.53 & 0.54 & 0.55 & 0.50\\
total power & 0.61 & 0.72 & 0.71 & 0.70 & 0.70\\
entropy & 0.29 & & 0.35 & 0.38 & 0.36\\
\hline
rel delta & 0.55 & 0.56 & 0.57 & 0.58 & 0.50 \\
rel theta & 0.79 & 0.78 & 0.78 & 0.77 & 0.75\\
rel alpha & 0.61 & 0.63 & 0.62 & 0.60 & 0.53\\
rel beta & 0.58 & 0.57 & 0.57 & 0.54 & 0.54\\
rel gamma & 0.42 & 0.51 & 0.53 & 0.51 & 0.48\\
\hline
\end{tabular}
\caption{Average correlation coefficients for qEEG measures over six sets of eyes-closed EEGs recorded at weekly intervals at the Cz electrode. The second column gives Pearson correlations for the original, untransformed quantities, the third column refers to quantities transformed according to commonly used transformations (log for absolute power and logit for relative power), the fourth column to quantities transformed using the Box-Cox method, and the fifth column to quantities transformed according to (7). The last column lists Spearman rank correlations. The third column is left blank for spectral entropy, which does not have a widely used normalizing transformation.}
\end{table}

It can be seen from Table~3 that transforming to normality improves test-retest reliability. Combining the results of Table~1 and Table~3 shows that the increase in correlation is not due only to the normality of the distributions, since log delta, log alpha, logit rel alpha, and logit rel gamma are significantly non-normal, but their test-retest reliabilities are comparable to those obtained after our transformation. A paired-samples \emph{t}-test was used to compare the original correlations for all spectral measures investigated with those obtained using (7). This yielded p=0.013, and the average correlation was increased from 0.59 to 0.64. The improvement tended to be most pronounced when the original variables were highly non-normal (e.g., delta and gamma power, and relative gamma power). Spearman correlations were slightly lower on average than Pearson correlations after transformation to the Gaussian (reduced from 0.64 to 0.61, p=0.0025). After optimization, results for the Box-Cox transformation were very similar to those obtained using (7) (p=0.25).

\section{Discussion}

We have derived a transformation of any given probability density as nearly as possible to the normal distribution. Results collectively equivalent to this transformation have been known in the technical statistics literature, but have made little or no impact in engineering or the behavioral, social, environmental, medical, physical or biological sciences, despite wide use of ad hoc transformations to achieve a similar outcome. 

The transformation appears to be particularly useful for quantities for which no transformation is as yet known to improve agreement with the Gaussian, and for quantities that deviate strongly from normality. It provides a systematic way of achieving normality, unlike the miscellany of ad hoc methods hitherto employed. It leads to better agreement with the Gaussian than any other method, including the Box-Cox (1964) transformation. Another advantage of the transformation is that the resulting distribution is symmetric, so that the mean coincides with the mode and the median, and there is a straightforward relation between quantiles and the standard deviation. This aspect, as well as the familiarity of the standard Gaussian form, is especially useful to facilitate comparing and interpreting differences between measured values. In addition, the method can be used to obtain distributions of any form, by taking the cumulative distribution function of the data, leading to a uniform distribution, and then applying the inverse CDF of the desired distribution.

In an illustrative example, the transformation was shown to improve the average test-retest reliability of qEEG parameters as measured by Pearson correlations, which confirms the findings of Dunlap et al. (1994). Pearson correlations after our tranformation were slightly higher on average than nonparametric Spearman rank correlations. Test-retest reliability using our transformation was also comparable to that obtained using the logarithm for absolute power measures, and the logit transformation for relative band powers, but the advantage of the transformation described here is that it can also be applied to quantities for which no transformation is known to approximate normality. The test-retest reliability of quantities transformed with the Box-Cox method was comparable to that obtained with our transformation. However, the advantage of our method over the Box-Cox method is that it is much easier to implement, and does not involve laborious optimization of parameter values. 

We stress that the transformation can be applied not only to EEG data, but to any non-normal data in neuroscience or other disciplines. There are many instances of such variables, and although the central limit theorem ensures that the average of many measurements of the same quantity will be approximately normally distributed, this argument does not hold in the study of test-retest reliability, which concerns separate measurements. Moreover, the distribution of values for different subjects does not necessarily tend to normal even if we compute averages of many different measurements for each subject. This is because measurements on different subjects constitute estimates of different quantities, whereas the central limit theorem governs the distribution of estimates of the same quantity.

As is the case for any transformation, applications should be chosen with care, since transforming to the Gaussian is not a panacea for problems arising from non-normality even if normality is an assumption of the relevant statistical test.

First, as always with empirical studies, one must minimize the presence of illegitimate observations by appropriate study design and careful collection of data. Moreover, most tests make additional assumptions besides normality, such as independence and identical distribution of the observations, constancy of error variance, and/or a simple (linear) relation to predictor variables. If other assumptions besides normality are important one may look for compromise transformations that give weight to these properties. 

For small sample sizes, the transformation may differ significantly from that in a repeat study, since the underlying distribution is incompletely represented in a small sample. This may complicate the comparison of results between studies. One solution is to use a single transformation based on the combined data sets, which is likely to render each partial distribution only approximately normal.

The power and specificity of a test for the comparison of means do not depend directly on normality of the samples after transformation. Rather, a suitable transformation should selectively increase the test statistic when the null hypothesis is false, and/or decrease it otherwise. Instead of using a normalizing transformation (which can help in some cases), one could apply bootstrapping techniques (see, e.g., Davison and Hinkley, 1997), or estimate improved confidence intervals for the \emph{t}-test based on the first three moments of the empirical distributions (Zhou and Dinh, 2005). However, these are much more complex undertakings.

Thus, while it is not a cure for all ills, the transformation described here has possible applications in a wide range of statistical problems where a distribution of normal (or any other particular) form is desirable.

\section{Acknowledgements}
This work was supported by the Australian Research Council, an Endeavour International Postgraduate Research Scholarship, and an International Postgraduate Award. We thank the Brain Resource International Database (under the auspices of the Brain Resource Company---www.brainresource.com) for the use of their data.

\newpage

\section{References}

\noindent Bellivier F, Golmard JL, Rietschel M, Schulze TG, Malafosse A, Preisig M, McKeon P, Mynett-Johnson L, Henry C, Leboyer M. Age at onset in bipolar affective disorder: further evidence for three subgroups. Am. J. Psychiat. 2003; 160: 999--1001.\\

\noindent Box GEP, Cox DR. An analysis of transformations. J. R. Stat. Soc. B., 1964; 26: 211--252.\\

\noindent Chen X, Tung Y. Investigation of polynomial normal transform. Struct. Saf., 2003; 25: 423--445.\\

\noindent Conover WJ. Practical Nonparametric Statistics. John Wiley \& Sons: New York, 1971: 309.\\

\noindent Davison AC, Hinkley DV. Bootstrap Methods and Their Application. Cambridge University Press: Cambridge, 1997.\\

\noindent Draper J. Properties of distributions resulting from certain simple transformations of the normal distribution. Biometrika, 1952; 39: 290--301.\\

\noindent Dunlap WP, Chen RS, Greer T. Skew reduces test-retest reliability. J. Appl. Psychol., 1994; 79: 310--313.\\

\noindent Gasser T, B{\"a}cher P, M{\"o}cks J. Transformations towards the normal distribution of broad band spectral parameters of the EEG. Electroencephalogr. Clin. Neurophysiol., 1982; 53: 119--124.\\

\noindent John ER, Ahn H, Prichep L, Trepetin M, Brown D, Kaye H. Developmental equations for the electroencephalogram. Science, 1980; 210: 1255--1258.\\

\noindent L{\'e}vy P. Th{\'e}orie de l'Addition des Variables Al{\'e}atoires. Gauthier-Villars: Paris, 1937: 71, 121.\\

\noindent Liu P-L, Der Kiureghian A. Multivariate distribution models with prescribed marginals and covariances. Prob. Eng. Mech., 1986; 1: 105--112.\\

\noindent Lloyd AJ, Ferrier IN, Barber R, Gholkar A, Young AH, O'Brien JT. Hippocampal volume change in depression: late- and early-onset illness compared. Br. J. Psychiat., 2004; 184: 488--495.\\

\noindent Napper GA, Pianta MJ, Kallonatis M. Localization of amino acid neurotransmitters following in vitro ischemia and anorexia in the rat retina. Visual Neurosci. 2001; 18: 413--427.\\

\noindent Oken BS, Chiappa KH. Short-term variablity in EEG frequency analysis. Electroencephalogr. Clin. Neurophysiol., 1988; 69: 191--198.\\

\noindent Rosenblatt M. Remarks on a multivariate transformation. Ann. Math. Stat., 1952; 23: 470--472.\\

\noindent Thode Jr. HC. Testing for Normality. Marcel Dekker: New York, 2002: 104.\\

\noindent Ultsch A. A Neural Network Learning Relative Distances. In Amari S-I, Giles CL, Gori M, Piuri V, editors. IJCNN 2000, Proceedings of the IEEE-INNS-ENNS International Joint Conference on Neural Networks. IEEE Computer Society Press: Los Alamitos, CA, 2000; 5: 553--558.\\

\noindent Zhou X-H, Dinh Ph. Nonparametric confidence intervals for the one- and two-sample problems. Biostatistics, 2005; 6: 187--200.

\end{document}